\documentclass{article}
\usepackage{spadre2008}
\usepackage{graphicx}
\frompage{000} \topage{000}                                              
\def\rightmark{Isotropization of quark-gluon plasma}
\def\leftmark{B. Zhang et al.}

\title{Pressure isotropization of
an equilibrating quark-gluon plasma}
\authors{
{Bin Zhang$^1$, Lie-Wen Chen$^2$ and Che-Ming Ko$^3$
}\\[2.812mm]
{\normalsize
\hspace*{-8pt}$^1$ Arkansas State University, \\
State University, AR 72467-0419, USA\\[0.2ex]
\hspace*{-8pt}$^2$ Shanghai Jiaotong University, \\
Shanghai 200240, China\\[0.2ex]
\hspace*{-8pt}$^3$ Texas A\&M University, \\
College Station, TX 77843-3366, USA
}} \abstract{Pressure isotropization 
of an equilibrating quark-gluon plasma produced in
relativistic heavy ion collisions is studied within the framework of
a multi-phase transport model (\textsc{AMPT}). The time evolution of
the bulk properties of the quark-gluon plasma is found to depend on
its expansion dynamics and hadronization scheme as well as the
scattering cross sections among quarks and gluons. It is further
found that the pressure isotropy of the
produced quark-gluon plasma can only be achieved temporarily,
indicating that there is only partial thermalization 
during the time evolution of the quark-gluon
plasma. } \keyword{Relativistic
Heavy Ion Collisions, Isotropization, Quark-Gluon Plasma}
\PACS{25.75.-q, 25.75.Nq, 24.10.Lx}

\begin{document}

\maketitle
\setcounter{page}{1}

\section{Introduction}\label{intro}
Both ideal hydrodynamics and non-equilibrium transport models can
describe many of the RHIC data on the collective dynamics of
produced matter. Although ideal hydrodynamics assumes local thermal
equilibrium, its underlying equations of motion can also be used for
the case when only local pressure isotropization is achieved
\cite{Heinz:2005zi}. It is thus of interest to know whether
pressure isotropization is achieved in transport models.
Using the \textsc{AMPT} model \cite{Zhang:2007rab}, we
have examined the pressure isotropization 
of the equilibrating quark-gluon plasma
produced in relativistic heavy ion collisions 
by focusing on the central
cell of the collisions where high density matter is produced. Only
contributions from active particles, i.e., 
those that still undergo scattering, are included. Particles
that have already frozen-out are not considered as participants in
the equilibration process and are excluded from the calculations. To
obtain the connection with the hydrodynamical approach, we have
considered the energy-momentum tensor of produced
matter and extracted from it the energy density and pressure. The
pressure isotropy is then characterized by the ratio of
the longitudinal pressure to the transverse pressure. In the
following, after a brief review of the \textsc{AMPT} model, we study
the evolution of the bulk properties and the pressure anisotropy of
the matter in the \textsc{AMPT} models and then give a brief
summary.

\section{The AMPT model}
The \textsc{AMPT} model
\cite{Zhang:1999bd,Lin:2000cx,Lin:2001yd,Lin:2004en} is a hybrid
model that uses different programs for different stages of
relativistic heavy ion collisions. The publicly available
\textsc{AMPT} code has two options: the default model and the string
melting model. While both models use initial conditions from the
Heavy Ion Jet INteraction Generator (\textsc{HIJING}) model
\cite{Wang:1991ht}, they treat differently the partonic stage and
its hadronization. The default model extracts mini-jet partons from
\textsc{HIJING} and uses Zhang's Parton Cascade (\textsc{ZPC})
\cite{Zhang:1997ej} to evolve the parton system. At partonic
freeze-out, these partons are reconnected with their parent strings
and then hadronize via the Lund string
fragmentation model \cite{Sjostrand:1993yb}. The produced hadrons
undergo further interactions in A
Relativistic Transport (\textsc{ART}) model \cite{Li:1995pr} until
hadronic freeze-out. In the string melting model, instead of using
the mini-jet partons, the partonic matter is formed by breaking up
all the hadrons from \textsc{HIJING} according to their valence
structures. The resultant quark-anti-quark plasma is again evolved
using the \textsc{ZPC} parton cascade. As there are no strings in
the system, quarks and anti-quarks hadronize by recombining with
each other according to a coalescence model. The produced hadrons
enter the \textsc{ART} model for final hadron evolution. It has been
found that the default model gives a good description of particle
spectra, but it underestimates their elliptic flows. The string
melting model, on the other hand, can describe only spectra below 1
GeV/$c$ and it gives, however, a good description of the anisotropic
flows of hadrons. 
Other observables, such as $J/\psi$ production
\cite{zhang1}, strange \cite{chen} and charm \cite{zhang2} flows,
and the two-pion correlation functions \cite{lin}, have also been
studied in the \textsc{AMPT} model.

\section{Pressure isotropization 
of quark-gluon plasma}

\begin{figure}[!htbp]
\centering
\includegraphics[scale=0.65]{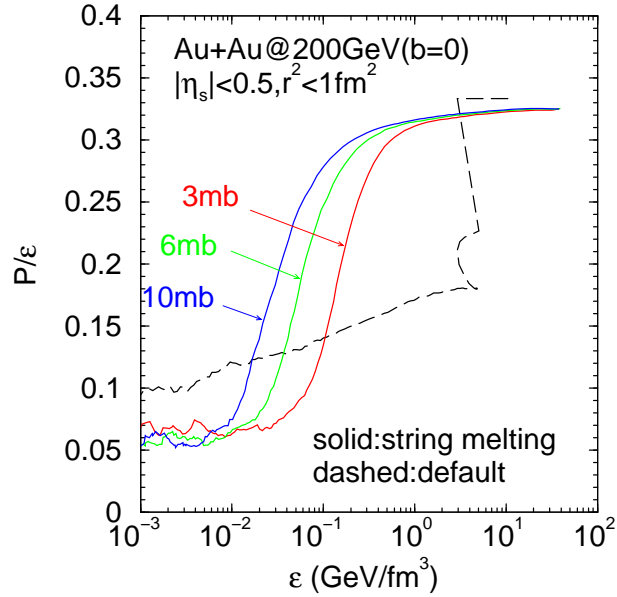}
\caption{\label{poe}Pressure to energy density ratio as a function
of energy density from the \textsc{AMPT} model.}
\end{figure}

The bulk properties of the hot and dense matter in the central cell
of central relativistic heavy ion collisions can be studied via its
pressure to energy density ratio as a function of energy density
(Fig.~\ref{poe}). This ratio gives the equation of state when the
matter under study is infinitely large and in equilibrium. The
central cell is specified by the space-time
rapidity, and the local rest frame momentum is used for the
calculation of energy density and momentum. The default model is
seen to hadronize at higher energy density (about 5 GeV/fm$^3$)
compared to the string melting model (hadronization is completed at
about two orders of magnitude below). This shows that these two
models serve as two limits: one is dominated by the hadron
description and the other by the parton description. A careful
examination further shows that in the default model, the system is
hadronic at 3 fm/$c$ while in the string melting model, the
formation of the hadronic matter is delayed to 13 fm/$c$.
Furthermore, the hadronization energy density becomes smaller as the
parton cross section increases. Unlike for a resonance gas in
equilibrium, the hadronic stage in the default model has a pressure
to energy density ratio that decreases as energy density decreases.
This happens when both the average hadron mass and average hadron
kinetic energy decrease as functions of time while the mass
to kinetic energy ratio increases. In other
words, heavier particles are left behind in the central cell,
leading to a reduced pressure to energy density ratio. The partonic
stage in the string melting model also has a decreasing pressure to
energy density ratio, and this is caused by the strange quarks that
are left behind. Consequently, the strange quark percentage in the
central cell increases with time.

\begin{figure}[!htbp]
\centering
\includegraphics[scale=0.65]{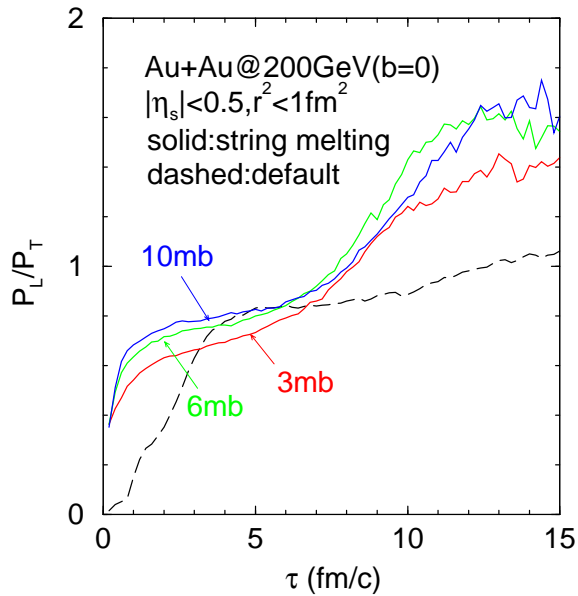}
\caption{\label{plopt}Evolution of pressure anisotropy.}
\end{figure}

Pressure isotropization can be characterized
by the time evolution of the longitudinal pressure (pressure along
the direction of the incoming nuclei (or beams)) to transverse
pressure ratio (Fig.~\ref{plopt}). In the \textsc{AMPT} model,
particle production follows the
inside-outside cascade picture of the Gyulassy-Wang model
\cite{Gyulassy:1993hr}. After two incoming nuclei pass through
each other, particles are first produced in the center of the space
between two receding nuclei and then produced near the
nuclei with higher longitudinal velocities. In the local rest frame
of produced matter, particles start with only transverse pressure.
The anisotropy then increases as thermalization proceeds. It is
clearly seen from Fig.~\ref{plopt} that in the string melting model
there is a faster increase of the pressure anisotropy as a function
of time. The pressure anisotropy crosses one at some time, but it
does not stay at one for any significant period of time. This
crossing is caused by the onset of transverse expansion as also seen
from the time evolution of the energy density. In the string melting
model, as the partonic cross section
becomes larger, the initial anisotropy
growth increases and its asymptotic value in
the longitudinal expansion stage is also higher. The case
with a 10 mb parton cross section can be
characterized by a relaxation time of about 0.5 fm/$c$ and an
asymptotic pressure anisotropy value of about 0.8. As full
pressure isotropization is only
achieved temporarily, the system can only reach
partial thermalization during the
collisions. Whether this can lead to a large deviation from
ideal hydrodynamical solutions is yet to be studied.

\section{Conclusions}\label{concl}

In summary, the pressure to energy density ratio in the default
\textsc{AMPT} model is much smaller than that in the string melting
\textsc{AMPT} model over a wide range of energy density. This also
depends on the partonic cross section. Both the longitudinal
expansion and the transverse expansion can affect the pressure
anisotropy. The hot and dense matter in the \textsc{AMPT} model does
not reach full pressure isotropization. This may have
implications on the difference in the description of the HBT radii
by ideal hydrodynamics and by transport models.

The default model and the string melting model in the current
\textsc{AMPT} model are two extreme descriptions of relativistic
heavy ion collisions. Improvements on the model, such as
hadronization at fixed time and inclusion of fragmentation
processes, can be made for a more coherent description of the
collisions. In addition, the effects of including parton number
changing processes, plasma instabilities, and mean fields in the
partonic matter are expected to lead to a better understanding of
the underlying physics in relativistic heavy ion collisions.

\section*{Acknowledgments}
We thank S. Bass, T. Renk, and U. Heinz for helpful discussions.
This work was supported by the U.S. National Science Foundation
under grant No.'s PHY-0554930 (B.Z.) and PHY-0457265, the Welch
Foundation under grant No. A-1358 (C.M.K.), the NNSF of China under
Grant Nos. 10575071 and 10675082, MOE of China under project
NCET-05-0392, Shanghai Rising-Star Program under Grant No.
06QA14024, and the SRF for ROCS, SEM of China (L.W.C.).

\vfill\eject
\end{document}